\documentclass[12pt]{article}
\usepackage{amsmath}
\usepackage{amssymb,latexsym}
\usepackage{multicol}
\usepackage{graphicx}
\usepackage{caption}
\usepackage{cite}
\usepackage{indentfirst}
\begin{document}
\title{Line-soliton and rational solutions to (2+1)-dimensional Boussinesq equation by Dbar-problem}
\author{ Junyi Zhu\thanks{Corresponding author.\newline\indent\quad{E-mail address:jyzhu@zzu.edu.cn}}\\
\small School of Mathematics and Statistics, Zhengzhou University,\\
\small Zhengzhou, Henan 450001, PR China}
\date{}
\maketitle
\begin{abstract}
We present a generalized (2+1)-dimensional Boussinesq equation, including two cases which are called the plus Boussinesq equation
and the minus one. To investigate these equations, we
apply the $\bar{\partial}$ approach to a coupled (2+1)-dimensional nonlinear equation, which reduces to the Boussinesq equation.
For the plus equation, we give the line-solitons and rational solutions, for the minus one, we give some freak solutions.
\end{abstract}
\section{Introduction}
The method based on the $\bar{\partial}$ (Dbar)-problem \cite{jpa21-L537,ip5-87,A-C1991,KBG1993} is a powerful tool to investigate the integrability of nonlinear PDEs ,
especially higher dimensional equations, and to find their explicit solutions, including solitons and rational solutions.
The departure from analyticity of a complex function can be measured by the Dbr derivative, a special case of this departure
is the jump condition for sectionally analytic function in the Riemann-Hilbert problem. However, if the associated eigenfunctions is analytic nowhere,
the Riemann-Hilbert approach fails, but the Dbar approach still works \cite{sam69-135}.

We consider a (2+1)-dimensional coupled  nonlinear equation \cite{apl69-15}
\begin{equation}\label{a1}
\begin{aligned}
u_t-2u_y+\frac{3}{2}\varepsilon u_{xx}-3\varepsilon v_x=0,\\
v_t-2v_y-u_y+\varepsilon u_{xxx}-uu_x-\frac{3}{2}\varepsilon v_{xx}=0,
\end{aligned}
\end{equation}
where $\varepsilon^2=\pm1$. The method we used in this paper is based on
the Dbar-problem in complex $k$ plane
\begin{equation}\label{a2}
\frac{\partial}{\partial\bar{k}}\chi(k,\bar{k})=\iint\chi(\lambda,\bar\lambda)R(\lambda,\bar\lambda;k,\bar{k}){\rm d}\lambda\wedge{\rm d}\bar\lambda,
\end{equation}
with canonical normalization $\chi\to1$ at $k\to\infty$. Here the domain of the integration is the complex $k$ plane, and will be omitted in the paper.

System (\ref{a1}) after elimination of $v(x,y,t)$ reduces to a generalized (2+1)-dimensional Boussinesq equation
\begin{equation}\label{a3}
u_{tt}-4u_{yt}+4u_{yy}-3\varepsilon u_{xy}+\frac{3}{4}\varepsilon^2u_{xxxx}-\frac{3}{2}\varepsilon(u^2)_{xx}=0,
\end{equation}
which can be reduces to the classical Boussinesq equation (see, e.g. \cite{amss3-217,cpam34-599,ip7-727,cpam35-567,jmp30-2201}).
The Boussinesq equation is integrable by the inverse problem method (see, \cite{jmp16-2301,pd6-51,pla79-264,cpam35-567,Z-M-N-P1984}), the Lax pair for this equation
was constructed in \cite{spj38-108}.
We now consider the following form of solution \cite{pd165-137,sam134-62}
\begin{equation}\label{a4}
u(x,y,t)=\kappa{\rm e}^{ikx-\varepsilon ik^3y-\varepsilon(2ik^3+\frac{3}{2}k^2)t},\quad \kappa\ll1, x,y,k\in{\mathbb{R}}, t>0.
\end{equation}
For $\varepsilon=1$, (\ref{a4}) is a small amplitude solution, which make the nonlinear term of (\ref{a3}) negligible.
In this case, the time frequency of (\ref{a4}) is complex, but we get the exponential growth at a rate of about ${\rm e}^{-\frac{3}{2}k^2t}$.
while, for $\varepsilon=i$, the amplitude of solution (\ref{a4}) is $\kappa{\rm e}^{-3k^3t}$, which is ill-posed for $k<0$.
In the following, we call the Boussinesq equation (\ref{a3}) with $\varepsilon=1$ the "plus" type, and $\varepsilon=i$ be "minus" type.

We note that Bogdanov and Zakharov investigated the continuous spectrum and soliton solutions for the two type Boussinesq equations
(named "plus" and "minus" Boussinesq equations) by using the Dbar-dressing method \cite{pd165-137}.
In that paper, the Boussinesq equation is a dimensional reduction in the framework of the KP hierarchy.
So the associated covariant derivative is related to that of KP equation.
In this paper, the generalized (2+1)-dimensional Boussinesq equation (\ref{a3}) is derived from different covariant derivative.

Although the two type equations have different properties, we present a useful method to discuss them at same time.
For the plus Boussinesq equation, we obtain line-soliton and rational solutions by choosing special degenerate kernel of the Dbar problem.
We note that the line-soliton solutions are complexiton solutions \cite{pla301-35,na70-4245,pa343-219,na63-e2461}, which show some periodic motions.
In addition, the rational solution can not be reduced to the lump solution for the plus Boussinesq equation.
For minus equation, we give some explicit solutions which show some strange appearance. Here we call them freak solutions.
It is remarked that the rational solutions for the plus Boussinesq equation also show some strange properties.

In \cite{apl69-15}, the author only studied the case of $\varepsilon=1$ of system (\ref{a1})
 by Darboux transformation and gave some periodic solutions. In the last part of this paper, we also extend the Darboux transformation
 to both cases of the Boussinesq equation, and obtain some new rational solution and freak solution, respectively.
 The present equation is different from the known (2+1)-dimensional Boussinesq equations studied in \cite{pla235-145,cma49-295,nar31-388}.

The paper is organized as follows. In Section 2, the Dbar dressing method is presented to derive the Lax pair of a coupled (2+1)-dimensional
nonlinear equation system, which can be reduced to the generalized (2+1)-dimensional Boussinesq equation.
In section 3, the explicit solutions, including line-solitons and rational solutions, are given by virtue of different choice of the kernel of the
Dbar problem. We give some discussions in the last section, and present some novel solutions by extending the method used in \cite{apl69-15}.

\section{Dbar dressing Method}
Since the time frequency of (\ref{a4}) is complex, we consider the wave number is $k$, in stead of $ik$, for convenience.
So, we introduce a set of operators $D_j$ as
\begin{equation}\label{b1}
D_1=\partial_x+k,\quad D_2=\partial_y+\varepsilon k^3,\quad D_3=\partial_t+\varepsilon(2k^3+\frac{3}{2}k^2),
\end{equation}
and the general kernel of the Dbar-problem is
\begin{equation}\label{b2}
R(\lambda,\bar\lambda;k,\bar{k})={\rm e}^{\theta(\lambda)}R_0(\lambda,\bar\lambda;k,\bar{k}){\rm e}^{-\theta(k)}.
\end{equation}
Here $R_0$ is an arbitrary function and
\begin{equation}\label{b3}
\theta(k)=kx+\varepsilon k^3y+\varepsilon(2k^3+\frac{3}{2}k^2)t.
\end{equation}

Suppose that the analytic function $\chi(x,y,t;k)$ has the following expansion
\begin{equation}\label{b4}
\chi(x,y,t;k)=1+\frac{1}{k}\chi^{(-1)}(x,y,t)+\frac{1}{k^2}\chi^{(-2)}(x,y,t)+\cdots,\quad k\to\infty.
\end{equation}
It is readily verified that the following two expressions have no singularity at $k\to\infty$
\begin{equation}\label{b5}
\begin{aligned}
D_2\chi-\varepsilon D_1^3\chi+uD_1\chi+v\chi=O(\frac{1}{k}),\\
D_3\chi-2\varepsilon D_1^3\chi-\frac{3}{2}\varepsilon D_1^2\chi+2uD_1\chi+(u+2v)\chi=O(\frac{1}{k}),
\end{aligned}
\end{equation}
where
\begin{equation}\label{b6}
u=3\varepsilon\partial_x\chi^{(-1)},\quad v=\varepsilon\left(3\partial_x^2\chi^{(-1)}-\frac{3}{2}\partial_x(\chi^{(-1)})^2+3\partial_x\chi^{(-2)}\right).
\end{equation}
Thus, according to the Liouville's theorem, we have two linear equations
\begin{equation}\label{b7}
L_1\chi:=D_2\chi-\varepsilon D_1^3\chi+uD_1\chi+v\chi=0,
\end{equation}
and
\begin{equation}\label{b8}
L_2\chi:=D_3\chi-2\varepsilon D_1^3\chi-\frac{3}{2}\varepsilon D_1^2\chi+2uD_1\chi+(u+2v)\chi=0.
\end{equation}
It is noted that the second linear equation takes another form
\begin{equation}\label{b8b}
\tilde{L}_2\chi:=D_3\chi-D_2\chi-\varepsilon D_1^3\chi-\frac{3}{2}\varepsilon D_1^2\chi+uD_1\chi+(u+v)\chi=0.
\end{equation}

Now, if one introduces new function $\psi(x,y,t;k)$ by the following transformation
\begin{equation}\label{b9}
\psi(x,y,t;k)=\chi(x,y,t;k){\rm e}^{\theta(x,y,t;k)},
\end{equation}
then the system (\ref{b7}) and (\ref{b8}) give rise to
\begin{equation}\label{b10}
\begin{aligned}
\psi_y-\varepsilon\psi_{xxx}+u\psi_x+v\psi=0,\\
\psi_t-2\varepsilon\psi_{xxx}-\frac{3}{2}\varepsilon\psi_{xx}+2u\psi_{x}+(u+2v)\psi=0.
\end{aligned}
\end{equation}
The compatibility condition $\psi_{ty}=\psi_{yt}$ of the system (\ref{b10}) imply the (2+1)-dimensional nonlinear equation (\ref{a1}).

\section{Solutions}
It is noted that the Dbar-problem (\ref{a2}) with the canonical normalization is equivalent to a integral equation
\begin{equation}\label{c1}
\chi(k,\bar{k})=1+\frac{1}{2\pi i}\iint\left\{\iint\chi(\mu,\bar\mu)R(\mu,\bar\mu;\lambda,\bar\lambda){\rm d}\mu\wedge{\rm }\bar\mu\right\}\frac{{\rm d}\lambda\wedge{\rm d}\bar\lambda}{\lambda-k}.
\end{equation}
In the case of the degenerate kernel
\begin{equation}\label{c11}
R(\mu,\bar\mu;\lambda,\bar\lambda)=\sum\limits_{j=1}^Nf_j(\mu,\bar\mu)g_j(\lambda,\bar\lambda),
\end{equation}
the linear integral equation (\ref{c1}) is reduced to the linear algebraic system.
Suppose that the solution of problem (\ref{c1}) has the asymptotic behaviors (\ref{b4}), then
\begin{equation}\label{c12}
\chi^{(-1)}=-\frac{1}{2\pi i}\sum\limits_{j,l=1}^N\xi_l(A^{-1})_{lj}\eta_j^0, \quad
\chi^{(-2)}=-\frac{1}{2\pi i}\sum\limits_{j,l=1}^N\xi_l(A^{-1})_{lj}\eta_j^1
\end{equation}
where
\begin{equation}\label{c13}
\xi_l=\iint f_l(\lambda,\bar\lambda){\rm d}\lambda\wedge{\rm d}\bar\lambda, \quad
\eta_j^m=\iint \lambda^mg_j(\lambda,\bar\lambda){\rm d}\lambda\wedge{\rm d}\bar\lambda, \quad (m=0,1),
\end{equation}
and
\begin{equation}\label{c14}
A_{lj}=\delta_{lj}+\frac{1}{2\pi i}\iint{\rm d}\mu\wedge{\rm }\bar\mu\iint\frac{{\rm d}\lambda\wedge{\rm d}\bar\lambda}{\lambda-\mu}g_l(\mu,\bar\mu)f_j(\lambda,\bar\lambda).
\end{equation}

\subsection{Line Solitons}
Existing the soliton solution is an important property for the integrable system. To derive the line solitons,
we consider functions $f_l$ and $g_j$ in the kernel (\ref{c11}) as
\begin{equation}\label{d1}
f_l(\mu,\bar\mu)=f_l{\rm e}^{\theta(\mu)}\delta(\mu-k_l), \quad g_j(\lambda,\bar\lambda)=g_j{\rm e}^{-\theta(\lambda)}\delta(\lambda-\tilde{k}_j),
\end{equation}
where $f_l, g_j$ and $k_l,\tilde{k}_j$ are constants. Here the function $\theta$ is defined by (\ref{b3}).
In this case, from (\ref{c13}) and (\ref{c14}), we have
\begin{equation}\label{d2}
\xi_l=-2if_l{\rm e}^{\theta(k_l)}, \quad \eta_j^m=-2ig_j\tilde{k}_j^m{\rm e}^{-\theta(\tilde{k}_j)},
\quad A_{lj}=\delta_{lj}+\frac{1}{2\pi i}\frac{\eta_l^0\xi_j}{k_j-\tilde{k}_l}.
\end{equation}

Now, the representations in (\ref{c12}) can be rewritten as the following form
\begin{equation}\label{d7}
\chi^{(-1)}=\frac{1}{2\pi i}\frac{\det A^{(a)}}{\det A}, \quad \chi^{(-2)}=\frac{1}{2\pi i}\frac{\det A^{(b)}}{\det A},
\end{equation}
the block matrix $A^{(a)}$ and $A^{(b)}$ are defined by
\begin{equation}\label{d8}
A^{(a)}=\left(\begin{matrix}
0&\hat\xi\\
\hat{\eta^0}^T&A
\end{matrix}\right), \quad A^{(b)}=\left(\begin{matrix}
0&\hat\xi\\
\hat{\eta^1}^T&A
\end{matrix}\right),
\end{equation}
where the row vectors $\hat\xi$ and $\hat{\eta^m}, (m=0,1)$ are defined as
\begin{equation}\label{d6}
\begin{aligned}
\hat\xi&=(\xi_1,\xi_2,\cdots,\xi_N), \quad &\hat{\eta^m}=(\eta_1^m,\eta_2^m,\cdots,\eta_N^m).
\end{aligned}
\end{equation}
Substitution (\ref{d7}) into (\ref{b6}), we have the explicit solution of the system (\ref{a1}).
We note that the representation $u=3\varepsilon\partial_x(\chi^{-1})$ will give the explicit solution of the generalized (2+1)-dimensional
Boussinesq equation (\ref{a3}). In the following, we will not to put stress on it.

Particularly, for $N=1$, we have
\begin{equation}\label{d3}
\chi^{(-1)}=-\frac{\exp(\theta(k_1)-\theta(\tilde{k}_1)+a)}{1+\frac{\exp(\theta(k_1)-\theta(\tilde{k}_1)+a)}{k1-\tilde{k}_1}}, \quad
\chi^{(-2)}=-\frac{\tilde{k}_1\exp(\theta(k_1)-\theta(\tilde{k}_1)+a)}{1+\frac{\exp(\theta(k_1)-\theta(\tilde{k}_1)+a)}{k1-\tilde{k}_1}},
\end{equation}
where $a=\ln(2if_1g_1/\pi)=\theta_0+i\varphi_0$. Furthermore, if let $k_1=k_R+ik_I, \tilde{k}_1=-k_R+ik_I, (k_R>0)$, we obtain
 one line-soliton solution
\begin{equation}\label{d11}
\begin{aligned}
u&=3\partial_x\chi^{(-1)}, \\
v&=3\partial_x^2\chi^{(-1)}-\frac{3}{2}\partial_x(\chi^{(-1)})^2+3\partial_x\chi^{(-2)},\\
\end{aligned}
\end{equation}
where
\begin{equation}\label{d12}
\begin{aligned}
\chi^{(-1)}&=-k_R\frac{{\rm e}^{\theta_1-\tau_1}+{\rm e}^{i\varphi_1}}{\cosh(\theta_1-\tau_1)+\cos{\varphi_1}}, \quad \tau_1=\ln(2k_R)\\
\chi^{(-2)}&=k_R(k_R-ik_I)\frac{{\rm e}^{\theta_1-\tau_1}+{\rm e}^{i\varphi_1}}{\cosh(\theta_1-\tau_1)+\cos{\varphi_1}}.
\end{aligned}
\end{equation}
Here, for $\varepsilon=1$,
\begin{equation}\label{d10}
\begin{aligned}
\theta_1&=2k_R[x+(k_R^2-3k_I^2)y+2(k_R^2-3k_I^2)t]+\theta_{0},\\
\varphi_1&=6k_Rk_It+\varphi_{0},
\end{aligned}
\end{equation}
and for $\varepsilon=i$,
\begin{equation}\label{d13}
\begin{aligned}
\theta_1&=2k_R(x-6k_It)+\theta_{0},\\
\varphi_1&=2k_R(k_R^2-3k_I^2)(y+2t)+\varphi_{0}.
\end{aligned}
\end{equation}

For the case of $\varepsilon=1$, that is solution (\ref{d11}) and (\ref{d12}) with (\ref{d10}), the frequency is two times the wave number in $y$ direction.
In this case, the one line-soliton shows a periodic motion (see Fig. 1), and the periodicity will be more
clear in the case $\varphi_1=(2n+1)\pi$ and $\theta_1-\tau_1=0$, which are removable singularities (see Fig. 2).

While, for the case of $\varepsilon=i$ and (\ref{d13}), the phase will play an important role in the wave motion, which give a freak wave (see Fig. 3).
 \begin{figure}[h]\
\begin{multicols}{3}
\includegraphics[width=5cm,height=4cm]{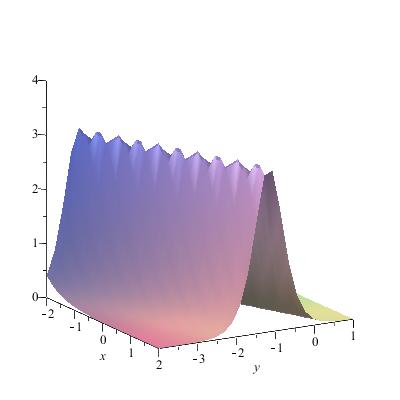}
\centerline{Abs(u)(t=0), $\varepsilon=1$}
\includegraphics[width=5cm,height=4cm]{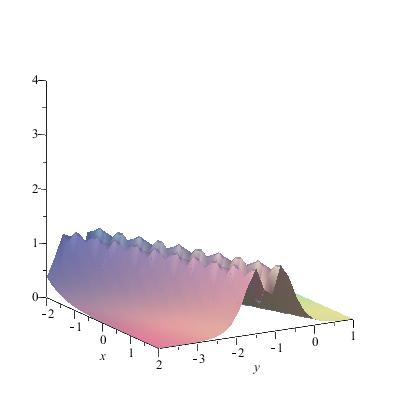}
\centerline{Abs(v)(t=0), $\varepsilon=1$}
\end{multicols}
\caption{\scriptsize one line-soliton $|u(x,y,t=0)|$ and $|v(x,y,t=0)|$ in \eqref{d11} with the parameters chosen as $k_R=k_I=1, \theta_0=\varphi_0=0$. }
\end{figure}

\begin{figure}[h]
\begin{multicols}{3}
\includegraphics[width=5cm,height=4cm]{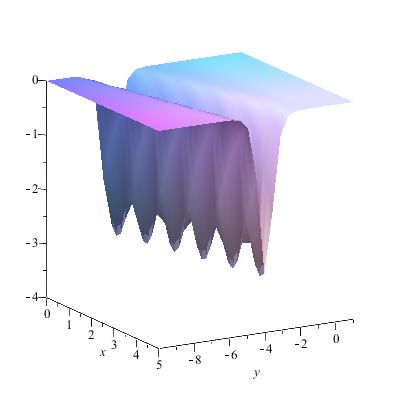}
\centerline{u(t=$\pi$), $\varepsilon=1$}
\includegraphics[width=5cm,height=4cm]{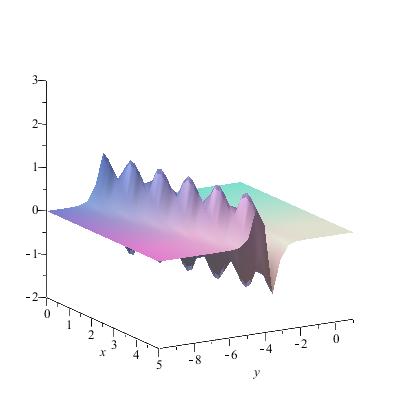}
\centerline{v(t=$\pi$), $\varepsilon=1$}
\end{multicols}
\caption{\scriptsize one line-soliton $u(x,y,t=\pi)$ and $v(x,y,t=\pi)$ in \eqref{d11} with the parameters chosen as $k_R=k_I=1, \theta_0=\varphi_0=0$.  }
\end{figure}

\begin{figure}[h]
\begin{multicols}{3}
\includegraphics[width=5cm,height=4cm]{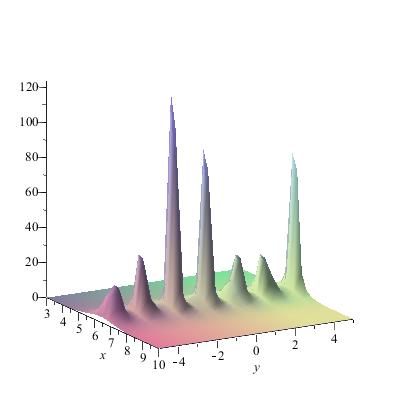}
\centerline{Abs(u)(t=$0$), $\varepsilon=i$}
\includegraphics[width=5cm,height=4cm]{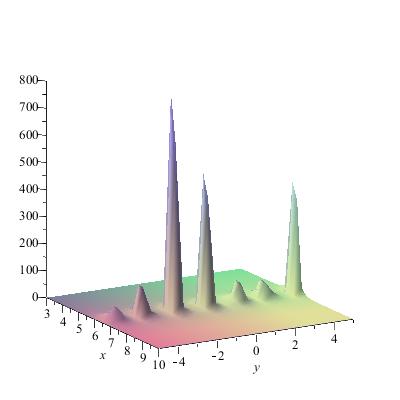}
\centerline{Abs(v)(t=$0$), $\varepsilon=i$}
\end{multicols}
\caption{\scriptsize one line-soliton $u(x,y,t=0)$ and $v(x,y,t=0)$ in \eqref{d11} with the parameters chosen as $k_R=k_I=1, \theta_0=\varphi_0=0$.  }
\end{figure}

For $N=2$ and $\varepsilon=1$, the figure of the two line-soliton (\ref{b6}),(\ref{d7}) and (\ref{d8}) with $k_1=1+0.2i,k_2=2+1.2i;\tilde{k}_1=-1+0.2i,\tilde{k}_2=-2+1.2i$,
$f_1=f_2=g_1=g_2=0.5i$ is shown in Fig. 4. In Fig. 5, it shows the two-freak wave for $N=2$ and $\varepsilon=i$.

\begin{figure}[h]
\begin{multicols}{3}
\includegraphics[width=5cm,height=4cm]{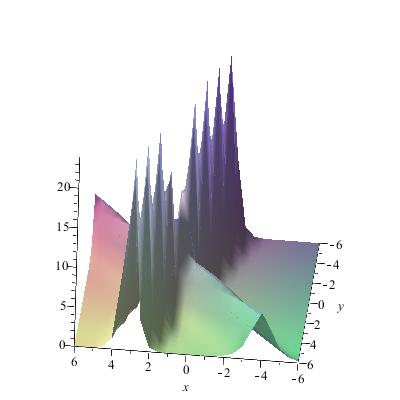}
\centerline{Abs(u)(t=0), $\varepsilon=1$}
\includegraphics[width=5cm,height=4cm]{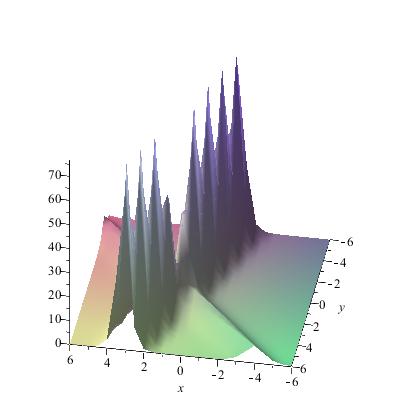}
\centerline{Abs(v)(t=0), $\varepsilon=1$}
\end{multicols}
\caption{\scriptsize two line-soliton $u(x,y,t=0)$ and $v(x,y,t=0)$ in \eqref{d11} with the parameters chosen as $k_1=1+0.2i,k_2=2+1.2i;\tilde{k}_1=-1+0.2i,\tilde{k}_2=-2+1.2i$ $ \theta_0=\varphi_0=0$.  }
\end{figure}
\begin{figure}[h]
\begin{multicols}{3}
\includegraphics[width=5cm,height=4cm]{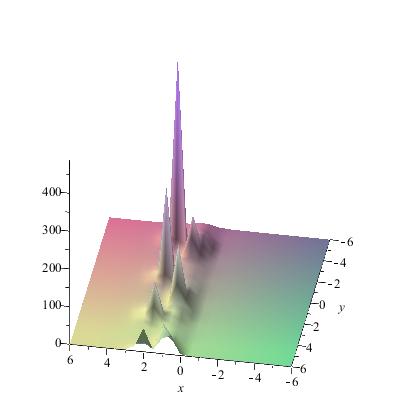}
\centerline{Abs(u)(t=0), $\varepsilon=i$}
\includegraphics[width=5cm,height=4cm]{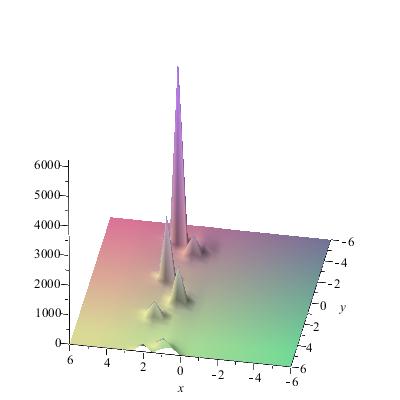}
\centerline{ Abs(v)(t=0), $\varepsilon=i$}
\end{multicols}
\caption{\scriptsize two line-soliton $u(x,y,t=0)$ and $v(x,y,t=0)$ in \eqref{d11} with the parameters chosen as $k_1=1+0.2i,k_2=2+1.2i;\tilde{k}_1=-1+0.2i,\tilde{k}_2=-2+1.2i$ $ \theta_0=\varphi_0=0$.  }
\end{figure}

\subsection{Rational solutions}
To obtain the rational solutions of the equations in (\ref{a1}), we choose the kernel as
\begin{equation}\label{c2}
R_0(\mu,\bar\mu;\lambda,\bar\lambda)=\sum\limits_{j=1}^{2N}f_j(\mu)\delta(\mu-k_j)g_j(\lambda)\delta(\lambda-k_j),
\end{equation}
where $f_j(\mu)$ and $g_j(\lambda)$ are smooth functions. Substituting $R={\rm e}^{\theta(\mu)}R_0{\rm e}^{-\theta(\lambda)}$ into (\ref{c1}), we have
a representation of $\chi(k,\bar{k})$
\begin{equation}\label{c3}
\chi(k,\bar{k})=1+\frac{2i}{\pi}\sum\limits_{j=1}^{2N}\frac{\chi_{j}f_jg_j}{k_j-k},
\end{equation}
where $\chi_j=\chi(k_j,\bar{k}_j), f_j=f_j(k_j), g_j=g_j(k_j)$. For convenience, we choose $f_i=\pi/(2i)$ and $g_j=1$,
then (\ref{c3}) reduces to
\begin{equation}\label{c4}
\chi(k,\bar{k})=1+\sum\limits_{j=1}^{2N}\frac{\chi_{j}}{k_j-k}.
\end{equation}
We note that $\chi_j$ can be obtained from (\ref{c1}) and (\ref{c2}) by letting $k=k_l$, that is
\[\chi_l=1+\sum\limits_{j\neq l}^{2N}\frac{\chi_j}{k_j-k_l}-\theta'(k_l)\chi_l, \quad l=1,2,\cdots,2N.\]
If one introduce the following matrices
\begin{equation}\label{c5}
\begin{aligned}
M=\left(\begin{matrix}
1+\theta'(k_1)&\frac{1}{k_1-k_2}&\cdots&\frac{1}{k_1-k_{2N}}\\
\frac{1}{k_2-k_1}&1+\theta'(k_2)&\cdots&\frac{1}{k_2-k_{2N}}\\
\cdots&\cdots&\cdots&\cdots&\\
\frac{1}{k_{2N}-k_1}&\frac{1}{k_{2N}-k_2}&\cdots&1+\theta'(k_{2N}),
\end{matrix}\right)\\
\check\chi=(\chi_1,\chi_2,\cdots,\chi_{2N}), \quad \check{E}=(1,1,\cdots,1)_{1\times2N}, \\
\theta'(k_j)=x+3\varepsilon k_j^2y+\varepsilon(6k_j^2+3k_j)t.
\end{aligned}
\end{equation}
then
\[\check\chi=\check{E}M^{-1}.\]

From (\ref{c3}), one find that $\chi(k,\bar{k})$ has the following asymptotic behavior
\[\chi(k,\bar{k})=1+\frac{1}{k}\chi^{(-1)}+\frac{1}{k^2}\chi^{(-2)}+\cdots, \quad, k\to\infty,\]
where
\begin{equation}\label{c6}
\chi^{(-1)}=-\check{E}M^{-1}\check{E}^T,\quad \chi^{(-2)}=-\check{k}M^{-1}\check{E}^T, \quad \check{k}=(k_1,k_2,\cdots,k_{2N}).
\end{equation}
We note that the representations of $\chi_{-1}$ and $\chi_{-2}$ take another forms
\begin{equation}\label{c7}
\chi^{(-1)}=\frac{\det M^{(a)}}{\det M}, \quad \chi^{(-2)}=\frac{\det M^{(b)}}{\det M},
\end{equation}
where $M^{(a)}$ and $M^{(b)}$ are $(2N+1)\times(2N+1)$ matrices
\begin{equation}\label{c8}
M^{(a)}=\left(\begin{matrix}
0&\check{E}\\
\check{E}^T&M
\end{matrix}\right), \quad M^{(b)}=\left(\begin{matrix}
0&\check{k}\\
\check{E}^T&M
\end{matrix}\right).
\end{equation}

Thus, we obtain the rational solution of the equation (\ref{a1})
\begin{equation}\label{c9}\begin{aligned}
u&=3\partial_x\left(\frac{\det M^{(a)}}{\det M}\right), \\
v&=3\partial_x^2\left(\frac{\det M^{(a)}}{\det M}\right)-\frac{3}{2}\partial_x\left(\frac{\det M^{(a)}}{\det M}\right)^2+3\partial_x\left(\frac{\det M^{(b)}}{\det M}\right).
\end{aligned}
\end{equation}
For $N=1$, we find
\[\begin{aligned}
\det M&=[1+\theta'(k_1)][1+\theta'(k_2)]+\frac{1}{(k_1-k_2)^2},\\
\det{M^{(a)}}&=-[2+\theta'(k_1)+\theta'(k_2)],\\
\det{M^{(b)}}&=1-[k_1+k_2+k_2\theta'(k_1)+k_1\theta'(k_2)],
\end{aligned}\]
where $\theta'(k_j)$ is defined in (\ref{c5}).

Particularly, if take $k_1=\xi+i\eta$, $k_2=\xi-i\eta$ and $\varepsilon=1$,  we get
\begin{equation}\label{c10}
u=3\partial_x(\chi^{(-1)}), \quad v=3\partial_x^2(\chi^{(-1)})-\frac{3}{2}\partial_x(\chi^{(-1)})^2+3\partial_x(\chi^{(-2)}),
\end{equation}
where
\[\begin{aligned}
\chi^{(-1)}=&\frac{-2(\tilde{x}+\xi\tilde{y})}{(\tilde{x}+\xi\tilde{y})^2+\eta^2\tilde{y}^2-\frac{1}{4\eta^2}},\\
\chi^{(-2)}=&\frac{1-2\xi\tilde{x}-2(\xi^2+\eta^2)\tilde{y}}{(\tilde{x}-\eta\tilde{y})^2+\xi^2\tilde{y}^2-\frac{1}{4\eta^2}},\\
\tilde{x}&=1+x-(\xi^2+\eta^2)(3y+6t),\\
 \tilde{y}&=2\xi(3y+6t)+3t.
\end{aligned}\]
The pictures of solution (\ref{c10}) at different time are shown in Fig. 6 and Fig. 7.
We note that, for the plus Boussinesq equation, the rational solution (\ref{c10}) is not a lump one.
As shown in Fig.7, the surface is strange. This picture is obtained at special time and in special region, shows some form of energy ejection.
\begin{figure}[h]
\begin{multicols}{3}
\includegraphics[width=5cm,height=4cm]{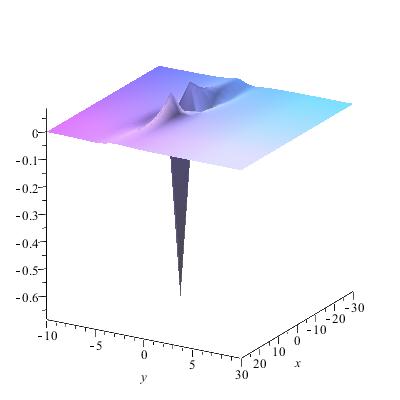}
\centerline{u(t=1), $\varepsilon=1$}
\includegraphics[width=5cm,height=4cm]{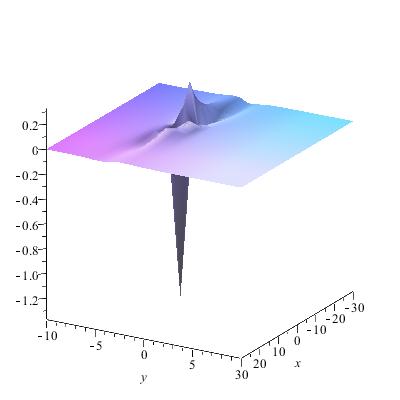}
\centerline{v(t=1), $\varepsilon=1$}
\end{multicols}
  \caption{\scriptsize $u(x,y,t=1)$ and $v(x,y,t=1)$ in \eqref{c10} with the parameters chosen as $\xi=2,\eta=1$.  }
\end{figure}

\begin{figure}[h]
\begin{multicols}{3}
\includegraphics[width=5cm,height=5cm]{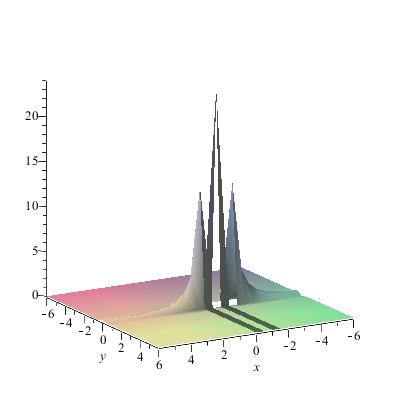}
\centerline{u(t=0), $\varepsilon=1$}
\includegraphics[width=5cm,height=5cm]{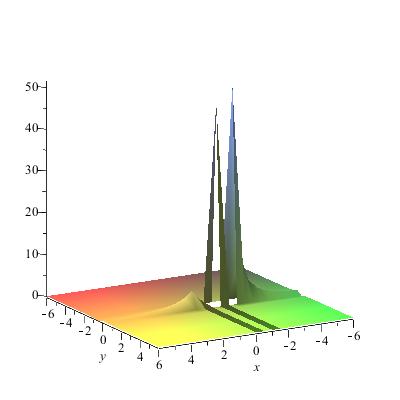}
\centerline{v(t=0), $\varepsilon=1$}
\includegraphics[width=5cm,height=5cm]{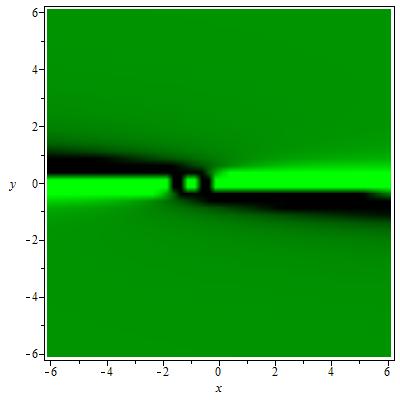}
\centerline{desityplot(u), $\varepsilon=1$}
\end{multicols}
  \caption{\scriptsize $u(x,y,t=0)$ and $v(x,y,t=0)$ in \eqref{c10} with the parameters chosen as $\xi=2,\eta=1$.  }
\end{figure}

\section{Discussions}
If we choose $f(\lambda,\bar\lambda)$ and $g(\lambda,\bar\lambda)$ in (\ref{c11}) as
$f(\lambda,\bar\lambda)=\tilde{f}(\lambda){\rm e}^{\theta(\lambda)}$ and $g(\lambda,\bar\lambda)=\tilde{g}(\lambda){\rm e}^{-\theta(\lambda)}$,
where $\theta(\lambda)$ is defined by (\ref{b3}), and assume that $\xi_j$ and $\eta_j^m, (m=0,1)$ exit in some region of $\lambda$ plane,
then $\xi_j$ and $\eta_j^m, (m=0,1)$ in (\ref{c13}) satisfy the system
\begin{equation}\label{e1}
\varepsilon\varphi_{xxx}=\varphi_y, \quad \varepsilon(2\varphi_{xxx}+\frac{3}{2}\varphi_{xx})=\varphi_t.
\end{equation}
However, if functions $\tilde{f}(\lambda)$ and $\tilde{g}(\lambda)$ are not the delta functions,
it is hard to give the explicit expression of $A_{lj}$ in (\ref{c14}). As a result, the solution of the generalized Boussinesq (\ref{a3}) or the system (\ref{a1})
will not be obtained by the Dbar-approach.

We note that an alternative method only starting from (\ref{e1}) can be introduced to obtained the solution of Boussinesq (\ref{a3}).
It is known that if $\varphi$ is a special solution of (\ref{e1}), then
\begin{equation}\label{e2}
u=-3\varepsilon(\ln \varphi)_{xx}, \quad v=-\varepsilon[3(\ln \varphi)_{xxx}+3(\ln \varphi)_{x}(\ln \varphi)_{xx}],
\end{equation}
will solve the system (\ref{a1}). We note that the case of $\varepsilon=1$ has been discussed in \cite{apl69-15}.
For example, a new particular solution of the system (\ref{a1}) can be given by choosing
\begin{equation}\label{e3}
\varphi=a(x^3+6\varepsilon y+9\varepsilon xt+12\varepsilon t)+bx+c,
\end{equation}
 where $a,b$ and $c$ are arbitrary constants. As shown in Fig. 8, there is a direction shift for the Boussinesq ($\varepsilon=1$) wave propagation.
\begin{figure}[h]
\begin{multicols}{3}
\includegraphics[width=5cm,height=5cm]{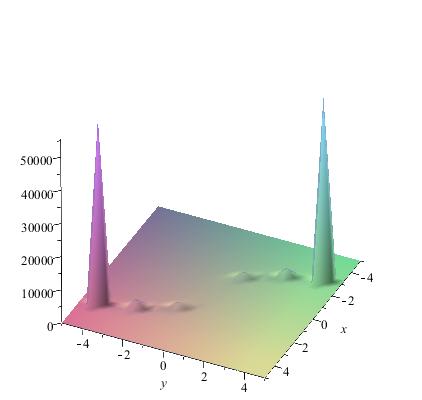}
\centerline{u(t=0), $\varepsilon=1$}
\includegraphics[width=5cm,height=5cm]{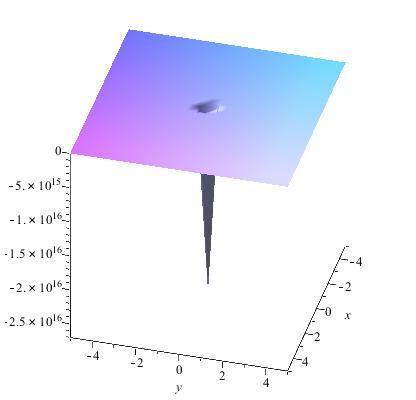}
\centerline{v(t=0), $\varepsilon=1$}
\end{multicols}
  \caption{\scriptsize $u(x,y,t=0)$ and $v(x,y,t=0)$ in (\ref{e2}) and (\ref{e3}) with $a=1,b=c=0$.  }
\end{figure}
\begin{figure}[h]
\begin{multicols}{3}
\includegraphics[width=5cm,height=5cm]{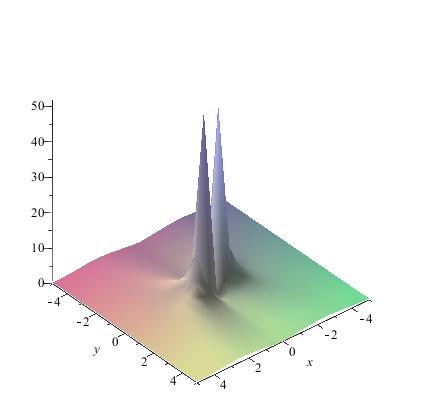}
\centerline{Abs(u)(t=0), $\varepsilon=i$}
\includegraphics[width=5cm,height=5cm]{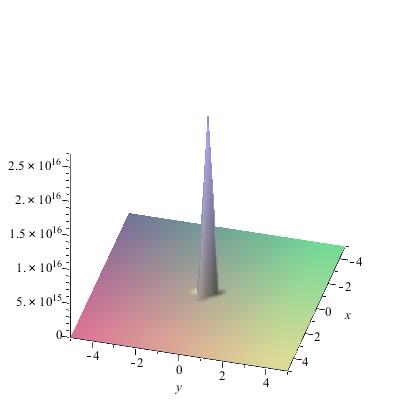}
\centerline{Abs(v)(t=0), $\varepsilon=i$}
\end{multicols}
  \caption{\scriptsize $Abs(u)(x,y,t=0)$ and $Abs(v)(x,y,t=0)$ in (\ref{e2}) and (\ref{e3}) with $a=1,b=c=0$.  }
\end{figure}

\section*{Acknowledgments}
Project 11471295 was supported by the National Natural Science Foundation of China.
%

\begin{thebibliography}{10}
\expandafter\ifx\csname url\endcsname\relax
  \def\url#1{\texttt{#1}}\fi
\expandafter\ifx\csname urlprefix\endcsname\relax\def\urlprefix{URL }\fi
\expandafter\ifx\csname href\endcsname\relax
  \def\href#1#2{#2} \def\path#1{#1}\fi

\bibitem{jpa21-L537}
L.~V. Bogdanov, S.~V. Manakov, The non-local partmacr problem and
  (2+1)-dimensional soliton equations, J. Phys. A: math. Gen. 21 (1988)
  L537--L544.

\bibitem{ip5-87}
R.~R. Beals, R.;~Coifman, Linear spectral problems, non-linear equations and
  the deltamacr-method, Inverse Problems 5 (1989) 87--130.

\bibitem{A-C1991}
M.~J. Ablowitz, P.~A. Clarkson, Solitons, Nonlinear Evolution Equations and
  Inverse Scattering, Cambridge University Press, Cambridge, 1991.

\bibitem{KBG1993}
B.~G. Konopelchenko, Solitons in Multidimensions---Inverse Spectral transform
  Method, Word Scientific, Singapore, 1993.

\bibitem{sam69-135}
M.~J. Ablowitz, D.~Bar~Yaacov, A.~S. Fokas, On the inverse scattering transform
  for the {K}adomtsev-{P}etviashvili equation, Stud. Appl. Math. 69 (1983)
  135--143.

\bibitem{apl69-15}
T.~Su, Explicit solutions for a modified 2+1-dimensional coupled {B}urgers
  equation by using {D}arboux transformation, Appl. Math. Lett. 69 (2017)
  15--21.

\bibitem{amss3-217}
H.~P. McKean, Boussinesq's equation as a {H}amiltonian system, Adv. Math. Supp.
  Studies 3 (1978) 217--226.

\bibitem{cpam34-599}
H.~P. McKean, {B}oussinesq's equation on the circle, Commun. Pure Appl. Math.
  34 (1981) 599--691.

\bibitem{ip7-727}
V.~A. Jurko, Solution of the {B}oussinesq equation on the half-line by the
  inverse problem method, Inverse Problems 7 (1991) 727--738.

\bibitem{cpam35-567}
C.~Deift, P.~Tomai, E.~Trubowitz, Inverse scattering and the {B}oussinesq
  equation, Commun. Pure Appl. Math. 35 (1982) 567--628.

\bibitem{jmp30-2201}
P.~A. Clarkson, M.~D. Kruskal, New similarity solutions of the {B}oussinesq
  equation, J. Math. Phys. 30 (1989) 2201--2213.

\bibitem{jmp16-2301}
M.~J. Ablowitz, R.~Haberman, Resonantly coupled nonlinear evolution equations,
  J. Math. Phys. 16 (1975) 2301--2305.

\bibitem{pd6-51}
P.~J. Caudrey, The inverse problem for a general {N}¡Á{N} spectral equation,
  Physica D 6 (1982) 51--66.

\bibitem{pla79-264}
P.~J. Caudrey, The inverse problem for the third order equation $u_{xxx} +
  q(x)u_x + r(x)u =-i\zeta^3u$, Phys. Lett. A 79 (1980) 264--266.

\bibitem{Z-M-N-P1984}
V.~E. Zakharov, S.~V. Manakov, S.~P. Novicov, L.~P. Pitaevsky, Theory of
  Solitons: The Inverse Scattering Method, Plenum Press, New York, 1984.

\bibitem{spj38-108}
V.~E. Zakharov, On stochastization of one-dimensional chains of nonlinear
  oscillations, Sov. Phys.¡ªJETP 38 (1974) 108--110.

\bibitem{pd165-137}
L.~V. Bogdanov, V.~E. Zakharov, The {B}oussinesq equation revisited, Physica D
  165 (2002) 137--162.

\bibitem{sam134-62}
A.~Himonas and D.~Mantzavinos, On the initial-boundary value problem for the linearized {B}oussinesq equation,
Stud. Appl. Math. 134 (2014) 62--100.

\bibitem{pla301-35}
W.~X. Ma, Complexiton solutions to the {K}orteweg-de {V}ries equation, Phys.
  Lett. A 301 (2002) 35--44.

\bibitem{na70-4245}
W.~X. Ma, A second wronskian formulation of the boussinesq equation, Nonlinear
  Anal. 70 (2009) 4245--4258.

\bibitem{pa343-219}
W.~X. Ma, K.~Maruno, Complexiton solutions of the {T}oda lattice equation,
  Physica A 343 (2004) 219--237.

\bibitem{na63-e2461}
W.~X. Ma, Complexiton solutions to integrable equations, Nonlinear Anal. 63
  (2005) e2461--e2471.

\bibitem{pla235-145}
M.~A. Allen, G.~Rowlands, On the transverse instabilities of solitary waves,
  Phys. Lett. A 235 (1997) 145--146.

\bibitem{cma49-295}
A.~M. Wazwaz, Variants of the two-dimensional {B}oussinesq equation with
  compactons, solitons, and periodic solutions, Comput. Math. Appl. 49 (2005)
  295--301.

\bibitem{nar31-388}
M.~J. Xua, S.~F. Tian, J.~M. Tua, T.~T. Zhang, B\"acklund transformation,
  infinite conservation laws and periodic wave solutions to a generalized
  (2+1)-dimensional {B}oussinesq equation, Nonlinear Anal. Real 31 (2018)
  388--408.

\end{thebibliography}

\end{document}